\documentclass{article}
\usepackage[margin=4cm]{geometry}
\usepackage{hyperref}

\usepackage{mathptmx}

\usepackage[document]{ragged2e} 
\usepackage[none]{hyphenat} 

\usepackage{amsmath, amssymb, graphicx, url, algorithm2e}

\usepackage{longtable}

\usepackage{authblk}

\usepackage{booktabs}
\usepackage[load-configurations=version-1]{siunitx} 
 
\usepackage[super]{natbib}
\setcitestyle{citesep={,}}
 
\usepackage[acronym,nomain,nonumberlist]{glossaries} 
\makeglossaries

\makeatletter
\def\set@curr@file#1{\def\@curr@file{#1}} 
\makeatother

\title{Why we do need Explainable AI for Healthcare}

\author[1, 2]{Giovanni Cinà, PhD}
\author[3]{Tabea Röber, MSc}
\author[3]{, Rob Goedhart, PhD}
\author[3]{Ilker Birbil, PhD}

\affil[1]{Institute of Logic, Language, and Computation, University of Amsterdam, NL}
\affil[2]{Pacmed, Amsterdam, NL}
\affil[3]{Business Analytics, Amsterdam Business School, University of Amsterdam, Amsterdam, NL} 

\usepackage{comment}

\begin{document}

\maketitle

\begin{abstract}
The recent spike in certified Artificial Intelligence (AI) tools for healthcare has renewed the debate around adoption of this technology. One thread of such debate concerns Explainable AI and its promise to render AI devices more transparent and trustworthy. A few voices active in the medical AI space have expressed concerns on the reliability of Explainable AI techniques, questioning their use and inclusion in guidelines and standards. Revisiting such criticisms, this article offers a balanced and comprehensive perspective on the utility of Explainable AI, focusing on the specificity of clinical applications of AI and placing them in the context of healthcare interventions. Against its detractors and despite valid concerns, we argue that the Explainable AI research program is still central to human-machine interaction and ultimately our main tool against loss of control, a danger that cannot be prevented by rigorous clinical validation alone.

\end{abstract}

\section{Introduction}

Along with the blooming of Artificial Intelligence (AI) and the accompanying increase in model complexity, there has been a surge of interest in explainable AI (XAI henceforth), namely AI that allows humans to understand its inner workings.\cite{doshi-velez2017, linardatos2020, Gilpin.2018, biran2017, doran2017} This interest is particularly keen in safety-critical domains such as healthcare, where it is perceived that XAI can engender trust, help monitoring bias, and facilitate AI development.\cite{doshi-velez2017, lipton2018} XAI has already shown to improve clinicians’ ability to diagnose and assess prognoses of diseases as well as assist with planning and resource allocation. For example, Letham and colleagues developed a stroke prediction model that matches the performance of the most accurate machine learning algorithms, while remaining as interpretable as conventional scoring methods used in clinical practice.\cite{letham2015}

Despite the enthusiasm, and a growing community of researchers devoting energy to XAI, there is currently no consensus on the reliability of XAI techniques and several researchers have cast serious doubts on whether XAI solutions should be incorporated into guidelines and standards, or even deployed at all.\cite{mccoy2021, Ghassemi.2021} 

Undoubtedly, there is an inherent tension between the desire for machines performing better than humans, and the desire for machines providing human-understandable explanations. Together with their super-human capacities -- such as the ability to juggle dozens or hundreds of factors -- the sub-symbolic character of statistical learning techniques contributes to rendering machine learning models opaque to humans.

There is a wide variety of techniques for XAI, and many categorizations have been proposed in the literature. Techniques can roughly be grouped into local vs. global, and model-specific vs. model-agnostic approaches. Local methods aim to explain model outputs for individual samples, while global methods focus on making models more explainable at an aggregate level. Model-specific methods are tailored to explain a specific type of model, while model-agnostic methods can be applied to a range of different models. Many of the well-known techniques yield post-hoc explanations, meaning that they generate explanations for already trained, so-called ‘black-box’, models. Alternatively, there exist approaches that are inherently explainable, also known as white-box models, such as decision trees and linear regression models. A detailed taxonomy is beyond the scope of this paper; for an extensive overview we refer the reader to existing reviews.\cite{carvalho2019, molnar_book2022, Ras.2022} 

What is important to stress is that, in order to overcome the opaqueness of black-box models, XAI techniques often bring in additional layers of complexity, such as:
\begin{itemize}
    \item explanations themselves can be inaccurate or deceiving,\cite{rudin_stop2019, Ghassemi.2021} 
    \item since an objective ground truth is lacking, it is hard to evaluate XAI techniques,\cite{zhou2021, yang2019} 
    \item trade-offs exist between explainability and other desiderata such as high performance.\cite{adadi2018, weld2019}
\end{itemize}

What role is left then for XAI in the medical realm, where adoption and trust of new technologies are notoriously complex? Given the wealth and heterogeneity of XAI techniques, as well as the doubts and limitations mentioned in the literature, the answer to this question is not straightforward.

In this Viewpoint we collect, review and appraise the criticisms that are leveled against XAI. We conclude that, despite several critical points of attention, there is no decisive argument to abandon the research for better XAI techniques, especially in the medical field, where a higher level of oversight is preferable.

\section{Criticisms to Explainable AI}

We begin by recollecting the main doubts raised about the usefulness and efficacy of XAI, in no particular order; each point is then expanded and discussed. The criticisms, along with the main counterpoints discussed below, are visually represented in Figure 1.
\begin{enumerate}
    \item XAI techniques may be incorrect or unreliable.
    \item Human users may fall prey to biases when interacting with XAI tools.
    \item There are no objective measures of explainability.
    \item XAI may work on an aggregate level, but not on an individual level.
    \item Evidence-based assessment is the proper way to evaluate medical (AI) interventions, not XAI.
    \item In the medical domain, we cannot trade performance for explainability.
    \item XAI is not suitable for all users.

\end{enumerate}

\begin{figure}[h]
    \centering
    \includegraphics[width=0.9\textwidth]{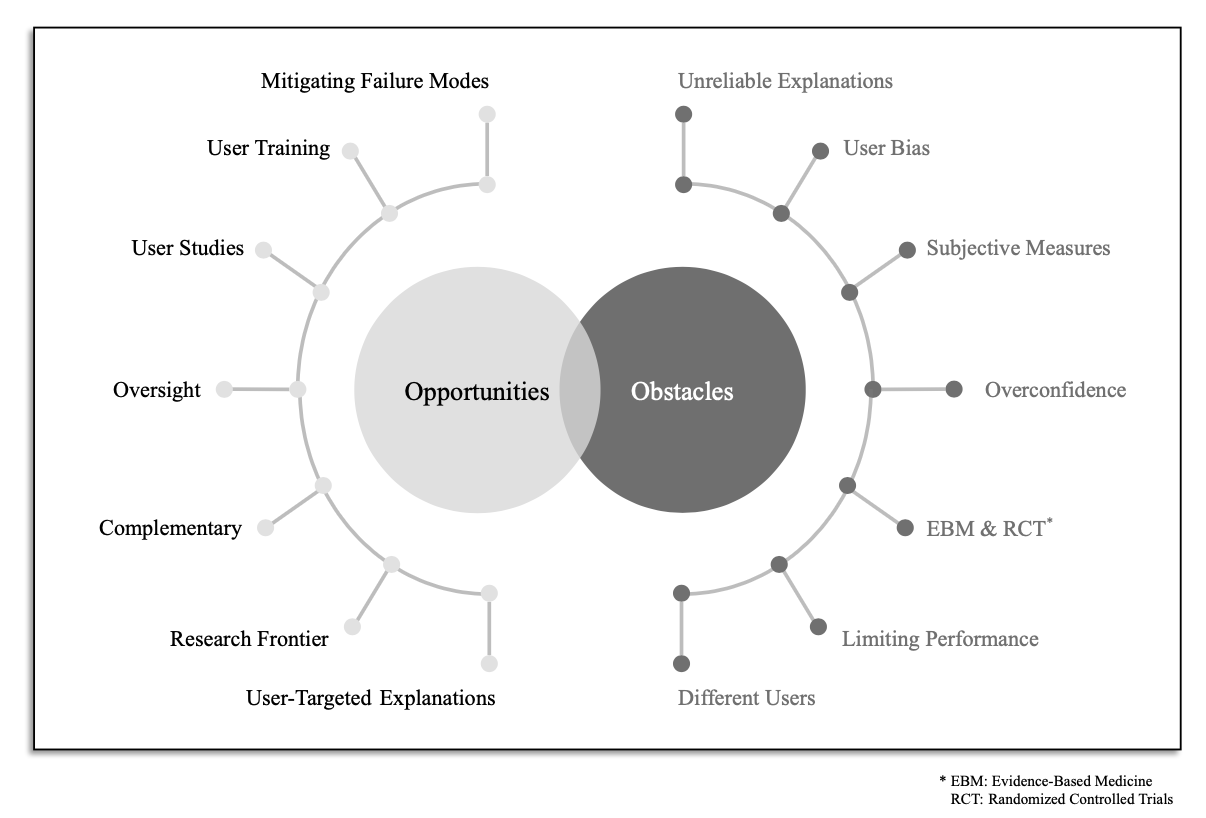}
    \caption{A visual summary of the criticisms and the related opportunities.}
    \label{fig:overview}
\end{figure}

\subsection*{XAI techniques may be incorrect or unreliable.}

\textbf{Criticism.} XAI techniques are sometimes faulty: they can be fooled with adversarial attacks or with noise.\cite{slack2019, heo2019}  Moreover, for the XAI techniques that approximate an original black-box model, there can be an additional discrepancy between the explanation and the real model behavior, resulting in unreliable explanations.\cite{rudin_stop2019}

\noindent
\textbf{Discussion.} While these points are true, the mitigating factor here is that every technology has failure modes. As long as we have awareness and know how to counter undesired effects, such techniques can still be useful. Furthermore, some such negative results can also be a pointer on how to improve XAI techniques; \textit{e.g.} adversarial attacks on SHAP (one of the techniques offering individual-level explanations) suggest that out-of-distribution samples should not be employed in approximate SHAP calculations.

\subsection*{Human users may fall prey to biases when interacting with XAI tools.}

\textbf{Criticism.} Users of XAI techniques struggle to understand what is a good explanation and often fall prey to biases –- e.g. preferring explanations that confirm their own beliefs, the so-called confirmation bias –- or misuse XAI tools.\cite{kaur2020, rosenfeld2021}

\noindent
\textbf{Discussion.} Such remarks are on point and should be addressed when designing interfaces of AI tools and training programs, but they do not constitute a sufficient reason to give up on XAI. It is very well documented that many researchers misunderstand and misuse $p$-values \cite{Wasserstein.2016}, but this is not a good argument to abandon significance testing. A similar point could be made about conventional medical procedures, where different kinds of bias have been documented. \cite{Schwartz.2014} The best solution for such problems is to have the users properly trained; notably, when it comes to medical devices the training of users is quite well regulated; \textit{e.g.} in Europe via the Medical Device Regulation and the ISO standards.\cite{mdr}

\subsection*{There are no objective measures of explainability.}

\textbf{Criticism.} There is no obvious ground truth for explainability, and therefore there is no straightforward way to assess how good an XAI technique really is.\cite{zhou2021, yang2019} 

\noindent
\textbf{Discussion.} This surely is a barrier, and more work is required to establish how the reliability of XAI tools can be tested; what is a good explanation, and what is useful to the user/developer/regulator are questions that have an intrinsic psychological nature and would benefit from more principled and widespread experimentation. Several of these concerns have been raised recently, \cite{Rudin.2022}  and substantial work has  already been done in investigating what constitutes an appropriate and useful explanation.\cite{slack2019, huysmans2011, miller2018, jacovi2022, krawiec2021} These efforts include eliciting users’ preferences with user studies, and deriving insights on the notion of explainability from the perspective of social sciences and related fields. Observing this need for an interdisciplinary structure, many scientists have recently participated in projects, workshops, discussion groups that involve colleagues working in diverse areas.

\subsection*{XAI may work on an aggregate level, but not on an individual level.}

\textbf{Criticism.} Ghassemi and colleagues distinguish XAI techniques used in the specific (individual level) case and in the general case, finding that they can be of use in the latter but not in the former.\cite{Ghassemi.2021} The reason is that users too often cannot decide if the explanation is sensible for an individual sample and have to base their judgment on an unwarranted intuition.

\noindent
\textbf{Discussion.} While this risk is concrete, it is important to realize that this criticism mostly applies to certain kinds of data. In images, single input features (i.e. pixels) have no intrinsic meaning; only certain patterns of feature activation have meaning. When a certain area of an image is highlighted we simply do not know if what we recognize (\textit{e.g.} the shape of a kidney) is the same as what the AI recognizes, because we have no access to the high-level representations of the machine. Hence, we cannot be sure whether the machine is recognizing the kidney as the salient part of the image at hand, or if there is some other spurious reason. There are however data types, like Electronic Health Records, where single input features have a very clear semantics. If, for instance, the patient's values of BMI and glucose are singled out as salient by an explainability technique, this may be enough for a clinical user to consider whether the AI's output is sensible. 

Thus in some applications, having XAI developed on an individual level in a way that allows users to ‘check’ the reason(s) behind the prediction, can be a way to find and escalate possible problems of the AI at an aggregate level. Especially when the data and the AI environment are of high complexity, we need oversight mechanisms to continuously monitor and detect unforeseen failure modes, and individual users can play a role in such monitoring processes. 
Furthermore, the explanations of AI predictions will typically be one of the pieces of information that clinicians take into account. In current practice, for example, physicians obtain results of different tests, which at times may even be potentially conflicting, \textit{e.g.} radiopaque lesion resembling infectious mass in a CT scan but lack of elevated infection markers. Such different factors are weighed in, each with its degree of uncertainty, to arrive at a decision. In this workflow, the input(s) of XAI techniques can be contrasted with other sources of information and be instrumental in i) achieving a deeper insight into the clinical case at hand and ii) decide whether to (not) trust the AI.


\subsection*{Evidence-based assessment is the proper way to evaluate medical interventions, not XAI.}

\textbf{Criticism.} According to this criticism, randomized controlled trials (RCTs) and evidence-based assessment are better ways to evaluate medical interventions than XAI. The primary goal is to have interventions that work, even if we do not understand why. Black-box solutions are used in medicine already, so we do not need XAI as long as we have established that a tool or intervention is effective. The desire for explainability is secondary.\cite{mccoy2021}

\noindent
\textbf{Discussion.} It is indisputable that XAI should not be conceived as a replacement of evidence-based instruments to validate model effectiveness. However, the criticism raises a false dichotomy: evidence-based evaluation tools and XAI can coexist, and we do not need to abandon one in favor of the other. We do not need to choose between a thorough evaluation and an explainable model: thorough evaluation should be a default (and it often is, given current regulations), and the need for explainability can be discussed separately. 

Furthermore, framing the discourse about explainability as a binary classification between primary and secondary requirements is needlessly reductionist. Surely there can be trade-offs with accuracy that we must be cognizant of (more on this later), but the benefit of explainability may pay off in other areas, \textit{e.g.} the discovery of biased outcomes for certain subgroups, or the disentangling of causal effects from spurious ones. As put forward by various governing bodies, explainability is one aspect within a set of requirements that are partially interdependent and should be balanced against each other. The pursuance of explainability will have different priority depending on the application. An AI tool to speed up reconstruction of images in the back-end might not need explainability, while a decision support tool to predict risk of complications might need it if the risks are to be discussed and assessed with a patient. A blanket statement declaring explainability ‘secondary’ does not help in determining which case is which. 

A final point should be made about validation of AI tools and evidence-based medicine (EBM). EBM can rely on a stable set of protocols that are established and widely used for the certification of new interventions. It should be noted however that, despite its indisputable importance, EBM is not a laser-precise discipline: e.g. there is a vast body of literature highlighting the fact that clinical trials are often not reflecting clinical practice, or have non-representative populations, or are incomparable with each other, or become quickly outdated.\cite{solomon2011}  Furthermore, as argued by several researchers,\cite{cartwright2012, Greenhalgh.2014, solomon2011, worrall2007} RCTs do not always control for the biases they are intended to control, they do not produce reliably generalizable knowledge, or they can create unnecessary constraints on clinical testing. In addition, obtaining statistically significant evidence is often extremely time-consuming and expensive. Finally, the success rate of clinical trials for medications is famously low.\cite{wong2019} Against this backdrop, we should be very careful to promote the initiation of many studies or trials testing the effectiveness of tools whose inner workings are entirely unknown, i.e. black boxes. Explainability is a precious tool not just for the end user, but also for the developer: the vast majority of models are discarded before reaching the test phase because they are found to be clinically implausible thanks to explainability techniques. In this sense, while we do not dispute the high position of RCTs in the hierarchy of evidence,\cite{djulbegovic2017} explainability is a complementary tool that can even help expedite or improve the effectiveness of RCTs and EBM as a whole. 


\subsection*{In the medical domain, we cannot trade performance for explainability.}

\textbf{Criticism. }The trade-off between explainability and performance may induce people to use simpler and less accurate models, rather than more effective ones. This is a problem, as we may be exchanging a further improvement in medical outcomes with a debatable increase in transparency. There are several medical interventions (\textit{e.g.} some drugs) for which a mechanistic explanation is not known, but they are nevertheless proven to be effective and widely used. If we are accepting of black-box interventions in healthcare, we can allow effective AI models to be opaque as well.\cite{mccoy2021}

\noindent
\textbf{Discussion.} It is indeed the case that some explainability techniques (or simpler, more interpretable models) can result in lower performance. Whether this is desirable is a decision that should be taken case-by-case. Moreover, the analogy with medications, and the fact that we are ‘fine’ with them being black-boxes, seems misguided. There are already different tools in medical practice for which different levels of understanding are required. For example, as opposed to drugs, radiography machines can only be operated by specialists. Hence for complex tools we do require users to have some knowledge of the inner workings of the device. The reason underlying this choice is that there are failure modes we want to be able to control.  
Moreover, the fact that now we do not require explanations for some interventions (as long as they are proven to work) does not mean that this is the desirable way forward.  On the contrary, in some cases this could be construed as a problem. Arguably, we should keep pushing research to understand why certain treatments are effective. Along this line of reasoning, if we can build a tool with an intrinsic ability to explain itself -- at a bearable cost in terms of trade-offs – we should definitely pursue it.

\begin{figure}[h]
    \centering
    \includegraphics[width=0.4\textwidth]{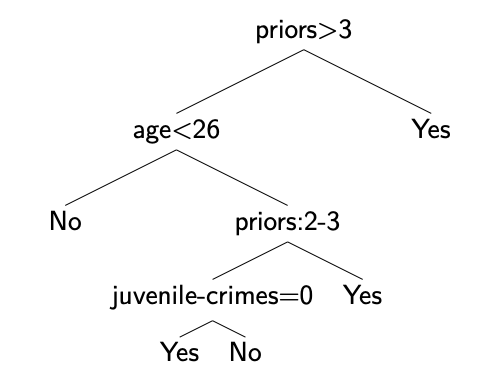}
    \caption{Example of an accurate white-box algorithm for the COMPAS recidivism problem (Figure 4 in \cite{Hu.2019}).}
    \label{fig:compas_example}
\end{figure}

Finally, a ‘simple' algorithm is not necessarily less accurate. When coupled with feature engineering or feature reduction, even those ‘simple' algorithms may achieve good performances. This criticism thus overlooks an important line of work on inherently interpretable methods that are aiming at matching the performances of more complex models.\cite{rudin_stop2019, letham2015} Figure \ref{fig:compas_example} showcases a simple model which is easy to understand and achieves high accuracy.


\subsection*{XAI is not suitable for all users.}

\textbf{Criticism.} XAI techniques can be used to support the work of different professionals interacting with an AI tool. Beside regular users, other roles include the developers, the medical experts, the regulators, and so on, each with different requirements and needs for explainability. Arguably, XAI techniques could be useful for trained professionals such as developers or auditors, but could be dangerous when shown to untrained users. 

\noindent
\textbf{Discussion.} This point has some overlap with points 2 and 4, but it is nonetheless important to stress that XAI techniques will almost never offer a one-size-fits-all solution. Different kinds of interaction with the machine will be required depending on the purpose of the participating human; it is highly likely that different techniques will answer different needs.\cite{Arrieta.2020} Current research extends to interactive machine learning to provide user-targeted explanations.\cite{sokol2020} As to the potential misuse or misunderstanding of XAI by simple users, the solution lies in a combination of robust methodology, proper training, and expectation management.

\section{Should we downsize our efforts in XAI for healthcare?}

We have reviewed some of the core criticisms offered in the literature against the widespread use of XAI techniques in medical contexts. Some of the qualms, as well as some of the counterpoints, are in fact not limited to healthcare applications and can be recast at a more general level. We should however be mindful of how the debate is influenced by the higher stakes present in safety-critical AI applications in the health realm.
It is worth remarking that demoting the usefulness of explainable medical AI means stifling the efforts to get human oversight over a new technology that is still in its infancy but can be deployed at scale very rapidly, potentially affecting a large swath of the population. XAI is not a silver bullet that will solve all of AI’s problems, and in particular it is not a substitute for rigorous evaluation of model performance. Nonetheless, XAI tools can be of value in several circumstances.
Human oversight over machines is still an important tenet, and the introduction of more black-boxes is not going to stem the loss of human control in a quickly digitizing world. It has been argued that augmentation of human capacities is preferable to automation,\cite{acemoglu2017} since the latter leads to deskilling as well as detachment and unemployment spikes. Augmentation presupposes the possibility of a human-machine synergy; in other words, a way for humans to relate and engage with AI. 

Different public bodies, including the European Union, are discussing regulations for AI products. While the field of XAI may not yet be at a level of maturity where XAI techniques are embedded into guidelines in the form of hard requirements, this does not entail that XAI techniques should be sidelined or disregarded. Said techniques can be employed to exercise an additional layer of control over AI products; as with all tools, practitioners should be aware of the potential failure modes. As for guidelines, it could be meaningful to allow the use of XAI techniques -- without enforcing a specific choice -- and require an argument of why a certain technique was chosen. 

We noted in the introduction the inherent tension in the request for explainability: while we want machines to perform at super-human level, we also want them to be understandable by humans. This is however just another case of competing objectives, not a good reason to toss one of the objectives out the window (for another example of competing objectives: saving healthcare costs and keeping people alive). The challenge is to find a trade-off that is satisfactory, a challenge that is extremely case-dependent. 
The question should be: while explainability may not be a requirement for healthcare interventions tout court, is AI a technology on which we think “extra” explainability is needed? This technology has peculiar aspects that set it apart, including the potential to be prejudiced against subgroups based on gender, ethnic background, and socioeconomic status,\cite{fitzgerald2017} the fact that it may lose reliability when the input data changes,\cite{ulmer_ood2020}  or that it may rely on spurious correlations found in the data.\cite{calude2017} For these reasons, coupled with the potential of rapid and impactful adoption, we argue that a higher level of oversight is preferable compared to regular interventions. 

\section{Conclusion}\label{sec:conclusion}

Despite its infancy, the research on XAI is growing very rapidly. Thanks to this growth, new techniques are tested extensively by many researchers in the field, and the duration to become a time-tested approach is much shorter than it has been in the past. Given the arguments we discussed, and the aforementioned possibility to rapidly hone useful techniques, we conclude that we should step up our efforts in XAI research for healthcare, as this research program is promising and addresses a core issue. Ultimately, we still want clinicians to be the mediators that can bridge the coldness of algorithms and scientific protocols with the human experience of the patient. If clinical users of medical AI cannot interact with machines, if their role becomes that of mere executors of interventions that are proven to be effective, the sensitive task of providing care could well become de-humanizing.









\end{document}